\documentclass[conference]{IEEEtran}
\IEEEoverridecommandlockouts
    \usepackage{comment}                                                       
\usepackage[autostyle]{csquotes}

 \usepackage[pdftex]{graphicx}
\usepackage{booktabs}
\usepackage{array}
\newcolumntype{P}[1]{>{\centering\arraybackslash}p{#1}}
\usepackage{cite}
\usepackage{amsmath,amssymb,amsfonts}
\usepackage{algorithmic}
\usepackage{graphicx}
\usepackage{textcomp}
\usepackage{xcolor}
\usepackage{bbding}
\def\BibTeX{{\rm B\kern-.05em{\sc i\kern-.025em b}\kern-.08em
    T\kern-.1667em\lower.7ex\hbox{E}\kern-.125emX}}

\begin{document}
%
\title{Demonstration Abstract: A Toolkit for Specifying Service Level Agreements for IoT applications
}

%
\author{
    \IEEEauthorblockN{Awatif Alqahtani\IEEEauthorrefmark{1}\IEEEauthorrefmark{2}\Envelope,  Pankesh Patel\IEEEauthorrefmark{3}, Ellis Solaiman\IEEEauthorrefmark{1}\Envelope,  Rajiv Ranjan\IEEEauthorrefmark{1}}
    \IEEEauthorblockA{\IEEEauthorrefmark{1}School of Computing Science, Newcastle University, Newcastle, UK
    \\\{a.alqahtani, ellis.solaiman, raj.ranjan\}@ncl.ac.uk}
    \IEEEauthorblockA{\IEEEauthorrefmark{2}Natural and Engineering, College of Applied Studies and Community Service, King Saud University, Riyadh, SA
    }
      \IEEEauthorblockA{\IEEEauthorrefmark{3}Fraunhofer CESE, College Park, Maryland,  USA
    \\ppatel@cese.fraunhofer.org}\\
    \textbf{Demo - Extended Version}
}




\maketitle

Today we see the use of the Internet of Things (IoT) in various application domains such as healthcare, smart homes, smart cars, and smart-x applications in smart cities. The number of applications based on IoT and cloud computing is projected to increase rapidly over the next few years. IoT-based services must meet the guaranteed levels of quality of service (QoS) to match users' expectations. Ensuring QoS through specifying the QoS constraints using Service Level Agreements (SLAs) is crucial. Therefore, as a first step toward SLA management, it is essential to provide an SLA specification in a machine-readable format. In this paper, we demonstrate a toolkit for creating SLA specifications for IoT applications. The toolkit is used to simplify the process of capturing the requirements of IoT applications. We present a demonstration of the toolkit using a Remote Health Monitoring Service (RHMS) usecase. The toolkit supports the following: (1) specifying the Service-Level Objectives (SLO) of an IoT application at the application level; (2) specifying the workflow activities of the IoT application; (3) mapping each activity to the required software and hardware resources and specifying the constraints of SLOs and other configuration- related metrics of the required hardware and software; and (4) creating the composed SLA in JSON format.



%
\IEEEpeerreviewmaketitle

\section{Introduction}

Service level agreement (SLA) in IoT involve many providers within the same architectural layer or between different architectural layers. For example, sensors that collect a patient's data are supplied by an actor that is different from the actor that provides a networking service in order to feed data to an IoT application, which in turn is provided by another actor. An actor is any participant in delivering an IoT application/service: it might be software, hardware or a human being. Each actor has responsibilities, abilities and requirements. To create an SLA, these different actors need to be considered, ensuring that each actor involved is delivering the required job within a certain level of Quality of Service (QoS), as well as receiving its requests within its QoS constraints. Therefore, this kind of dependability needs to be reflected and captured, and currently available SLA specification languages do not reflect this need~\cite{101}~\cite{102}.
 Several key challenges need to be considered to enable a shift from previous SLA specification languages to an SLA specification language for IoT applications:
\begin{itemize}
\item \textit{Challenge 1: Multi-layer nature of IoT Applications.} It is challenging to achieve a high level of certainty regarding the quality of a service, especially when the service is composed of/depends on other services \cite{80}.  Cloud based IoT applications are typically composed of a number of services in order to be able to perform the required functionality. These services can be deployed, or can consist of other services which are deployed, on one of the following computing resource layers:
\begin{itemize}
\item IoT device layer: Ð	Provides smart devices with the ability to sense and generate a large amount of data at different data speeds.
\item Edge resource layer: Provides the intelligence (computation) on the Edge devices to improve performance by reducing unnecessary data transferring to Cloud data-centers.
\item Cloud computing Layer: Offers a pool of configurable resources (hardware/software) that are available on demand \cite{18}, allowing users to submit jobs to service providers on the basis of pay-per-use. Cloud computing also offers advanced technologies for ingesting, analyzing and storing data  \cite{10lit}.
\end{itemize}
\item \textit{Challenge 2: Heterogeneity:} 
\begin{itemize}
\item Heterogeneity of  workflow activity: 	Different models of IoT applications have different stacks of essential interdependent services. Therefore, there is a heterogeneity issue in both sides, hardware and software.

\begin{itemize}
\item \textit{Heterogeneity related to hardware components:} Turning the light off in an auto-building application can be run on IoT and Edge layers, while comparing patients' old records with current recorded data in RHMS requires performing some tasks on IoT, Edge and Cloud layers. 
\item \textit{Heterogeneity related to software components: } Some applications require a certain type of data analysis programming model, such as applying data ingestion and stream processing to monitor a patient's health remotely. While  other applications, for example, in computing statistics of a particular vehicle for a month-long period, require ingestion, stream processing and batch processing data analysis programming models \cite{ozm}.
\end{itemize}
\item Heterogeneity of the key QoS metrics across layers \cite{31}:
The QoS requirements will depend on the required service. \textit{precision} when collecting data; for example, is a QoS requirement at IoT devices layer, while \textit{response time} is one of the key QoS requirements for  batch processing as programming model run on Cloud layer. 
\item Heterogeneity of application requirements \cite{31}:
At the application level, the QoS requirements vary from one application to another; for example, \textit{responsiveness} is a key requirement for health monitoring applications while  \textit{energy efficiency} is an essential QoS requirement for building automation applications.\\
\end{itemize}
\end{itemize}

In this paper, we introduce a toolkit for creating SLA for IoT applications. The toolkit supports the following: (1) specifying the Service-Level Objectives (SLO) of an IoT application at the application level; (2) specifying the workflow activities of the IoT application; (3) mapping each activity to the required software and hardware resources and specifying the constraints of SLOs and other configuration-related metrics of the required hardware and software; and (4) creating the composed SLA in JSON format. It is designed to be used by people who are interested in requesting/offering SLA of IoT applications with basic technical knowledge of IoT, Edge and Cloud technologies (e.g., IoT administrators).\\

\textbf{Outline}: The remainder of this paper is organized as follows: 
in Section \ref{sec:others}, we describe the state-of-the-art; in Section~\ref{sec:sys}, we present an overview of the system (design goals and system architecture); Section~\ref{sec:demonstration}  provides the demonstration plan by walking through a use-case. We conclude and present future work in Section~\ref{sec:conc}. 
\section{State-Of-The-Art}\label{sec:others}
There are a number of different approaches to specifying an SLA, from employing natural language or a formal language for the purpose of analyzing SLA properties, to utilizing XML documents in an effort to standardize SLAs to increase SLA interoperability between the service consumer and the service provider \cite{8}. For example, Keller and Ludwig provided an XML framework to express SLAs for Web Services (WSLA), which is considered to be a starting point as others have extended their approach  \cite{9}. Some efforts in SLA specifications have been made for the Cloud computing paradigm, such as in CSLA\cite{25}, SLAC\cite{19}. However,
SLAC\cite{19} has considered only an IaaS layer, while  CSLA\cite{25}
has considered all three Cloud delivery models (IaaS, PaaS, SaaS).  CSLA\cite{25} allows for specifying a PaaS layer as a whole layer, which means it does not consider specifying the QoS requirements of each data analysis programming model within that layer. Moreover, service providers typically provide inflexible take-it-or-leave-it SLAs, which disregard the fact that service consumers have varying requirements depending on their needs, budget and preferences (e.g., SLAs of AWS services and Microsoft Azure). Therefore, we aimed in our specification to consider the most common IoT application tiers/services, including data sources (e.g., sensors and RFID tags), programming models (e.g., stream processing and batch processing) and computing resources (e.g., Edge resources and Cloud resources) to allow users (e.g., IoT administrators) to specify their preferences.

\section{System Overview}\label{sec:sys}
In this section, we present the design goals and the architecture of the toolkit.

\subsection{Design Goals}
SLA creation is an important and critical step considering the fact that SLA-based service discovery, negotiation, monitoring, management and resource allocation rely on what has been specified within the SLA. As a result, we have developed a toolkit that enables service consumers to specify their QoS requirements and express them as service level objectives (SLOs), as well as specifying some configuration-related metrics for each software/hardware component of the system. We have considered the following features as \textit{design goals} of the toolkit:
\begin{itemize}
\item Expressiveness: We aim to provide a rich list of domain specific vocabularies to allow fine-grained SLA specification.

\item 	Generality: We aim to consider common components or layers of IoT architecture (IoT, Edge and Cloud).

\item 	Extendibility: ¥	The tool is to some extent extendable, because it has been designed to allow anyone who is interested in customising/enhancing the SLA according to his/her application-specific need to add or delete activity/metrics without changing the programming code. It is possible to add/delete/change activity/metrics using an attached Excel file, and these changes can be reflected dynamically. The Excel file preserves the schema of SLA components (e.g., workflow activities and their related software and hardware requirements).
\item 	Simplicity: Providing a GUI enables users to specify their requirements without needing prior knowledge of a machine-readable language such as JSON or XML. Furthermore, the tool allows users to specify an SLA in the same data-flow as their application, by allowing users to specify the workflow activity of their application first and then specify the requirements in the same flow of occurrences as the selected activities.
\end{itemize}
\subsection{System Architecture} 

\begin{figure*}
\centering
\includegraphics[width=0.8\linewidth, height=2in]{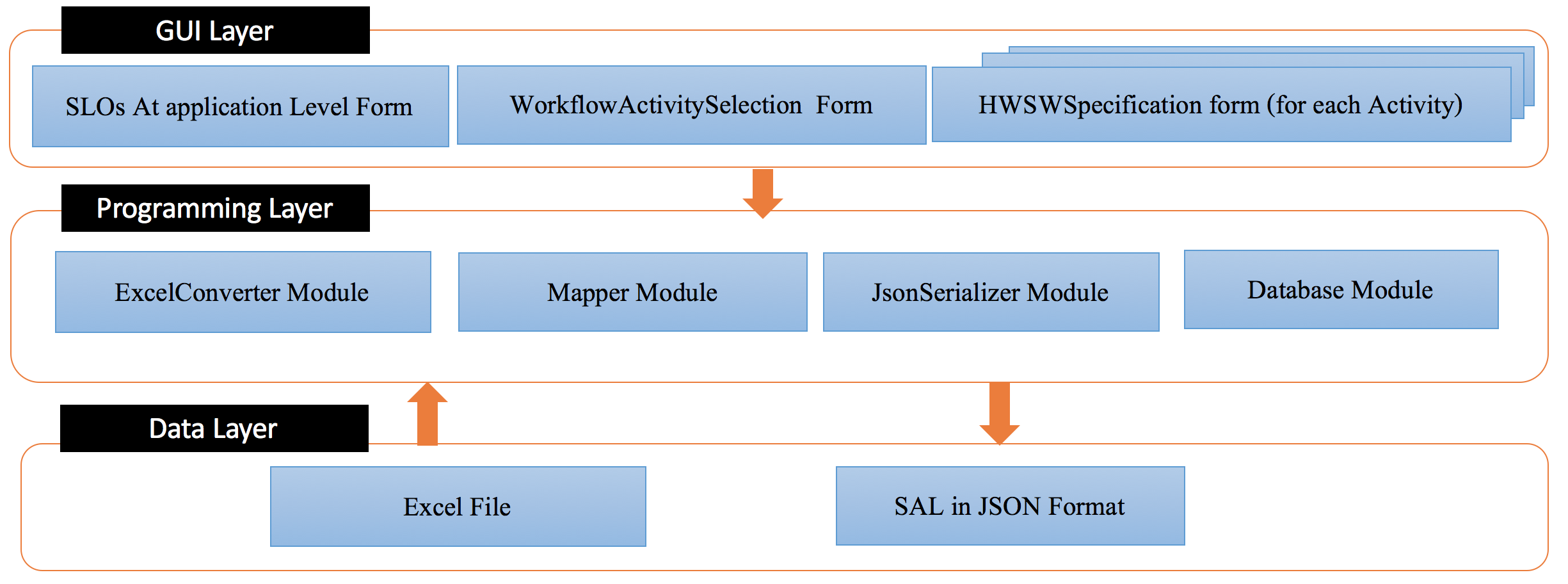}
\caption{Layered Architecture}
\label{fig:sys}
\end{figure*}

The abstracted design and architecture of the toolkit is depicted in  Figure \ref{fig:sys}. The overall architecture comprises three basic layers: 
\begin{enumerate}
\item GUI Layer, which includes the user interface components. It displays user interface components as a sequence of forms that guide the user through well-defined steps.
  \item Programming Layer, which encapsulates the programming modules in order to serve the GUI layer by providing the required functionalities.
  \item Data Layer, which encapsulates the required data as an input to the tool or output of the tool. It includes:

\begin{itemize}
\item 	An Excel file as an input, which provides data that describes the SLO and configuration metrics related to the software and hardware requirements of each activity. The Excel file has new vocabularies which provide fine-grained details regarding the associated software and hardware components of each workflow activity.
 
\item A JSON document as an output, which represents the SLA document. The specification is related to the SLOs at the application level followed by a specification related to each activity, which includes the SLOs and configuration related metrics required for the programming model (e.g., stream processing) and deployment layer (e.g., Cloud layer).
 
\end{itemize}
\end{enumerate}

\begin{figure}
\centering
\includegraphics[width=3.5in,height=0.75in]{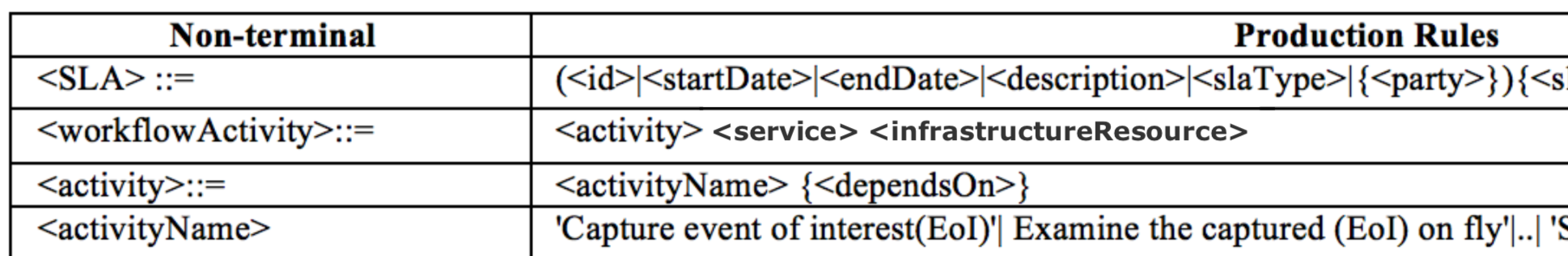}
\caption{A snippet of the proposed SLA grammar which preserves the structure of the SLA}
\label{fig:gr}
\end{figure}

For example, when the user starts the program, the  \textit{AppSLOs}  form is displayed to allow the user to specify the basic details concerning: type of IoT application (e.g., Remote Health Monitoring); preferable duration time to start the agreement (e.g., start date and end date); and SLOs at application level (e.g., end-to-end response time). After that, the user can select the workflow activities of an IoT application using the \textit{WorkflowActivitySelection} form. The form works in conjunction with the  \textit{ExcelConverter} module, which retrieves the names of the Excel file sheets as workflow activities to allow users to select which one is included in their application's workflow. Once the user has selected the workflow activities of the application, the \textit{Mapper} module maps each activity to the corresponding deployment layer as well as to the required programming model, if needed. For example,  \enquote{Large scale real-time analysis} is mapped to \enquote{Stream processing} as the programming model and is deployed on Cloud layer. The \textit{ExcelConverter} displays the related-metrics of SLOs and configuration for both the deployment layer and programming model of each activity based on the predefined schema within the Excel file. After users specify their constraints in both SLOs and configuration requirements for all selected activities, the \textit{Grammar} module ensures that an SLA object is created in such a way that it complies with the predefined grammar structure (Figure \ref{fig:gr}). The \textit{JsonSerializer} then comes into play, which is a software component to serialise the SLA object to be an SLA document in a JSON format.   The \textit{Database}  module stores the SLA object in MonogoDB database.

\section{Demonstration}\label{sec:demonstration}
The goal of our demonstration is to show how to create an SLA for an IoT application using a GUI-based toolkit. The tool considers domain-specific metrics for SLA specification purposes. Requirement of the Demo:\begin{itemize}
\item JRE (Java Runtime Environment): To run Java JAR file.
\item Microsoft Excel: To view the content of the Excel file.
\item Expected duration of the demo is 10 minutes as a demo format.
\end{itemize}

\subsection{Remote Health Monitoring Service (RHMS) Use-case}
To illustrate the usefulness of the toolkit, consider a Remote Healthcare Monitoring Service (RHMS) where patients wear sensors and accelerometers to measure their heart rate and sugar levels (capture data with high accuracy guarantees), reminding them of the time to take medications and detecting abnormal activity such as falling down (analyse incoming data on fly under low latency constraints). Data transferring from sensors to the Cloud layer requires 100\% network connectivity. Patients can register in RHMS and pay for the service to monitor their health remotely and alert their carers and doctors if their health is in a critical condition. Subscribed patients are looking for a service that can satisfy the following high-level requirement: detecting abnormal activity (such as falling down) within x milliseconds; ambulance, carers and doctors to be contacted within y minutes \cite{40}. Adherence to SLA's constraints of RHMS is a critical process. For example, if there was a delay in the network, it would lead to a late response at the front-end which exceeds what the consumer was expecting. From the above scenario, it can be seen that to achieve the high-level requirement, many nested-dependent QoSs should be considered. Therefore, the toolkit defines a multilayer specification with new vocabulary to allow for specifying the constraints for all back-end services, which cooperate to deliver the front-end service. Therefore, it is important that both service consumers and providers state the QoS requirements for each required service within the agreement clauses of the SLA.
\subsection{Demonstration Scenario}
Our demonstration would involve the following steps:
\begin{itemize}
\item \textit{Specify the SLOs of the Application at the application level:}
The tool will display a predefined list of possible SLOs as a check list. Users can check the SLOs that they are interested in and specify the priority level (high, low, or  normal) as well as the threshold value of the QoS metric of the SLO (Figure 3-a). In RHMS, users can quantify the SLOs by specifying the acceptable threshold value of the related Quality of Service metrics. For example, the objective of minimizing response time can be specified by selecting the preferred value for each of the following attributes related to the minimizing response time objective: priority (e.g., high); required level (e.g., less than); value (e.g., 60); unit (e.g., seconds).

\begin{figure} 
    \centering
       \includegraphics[width=3.5in,height=2.5in]{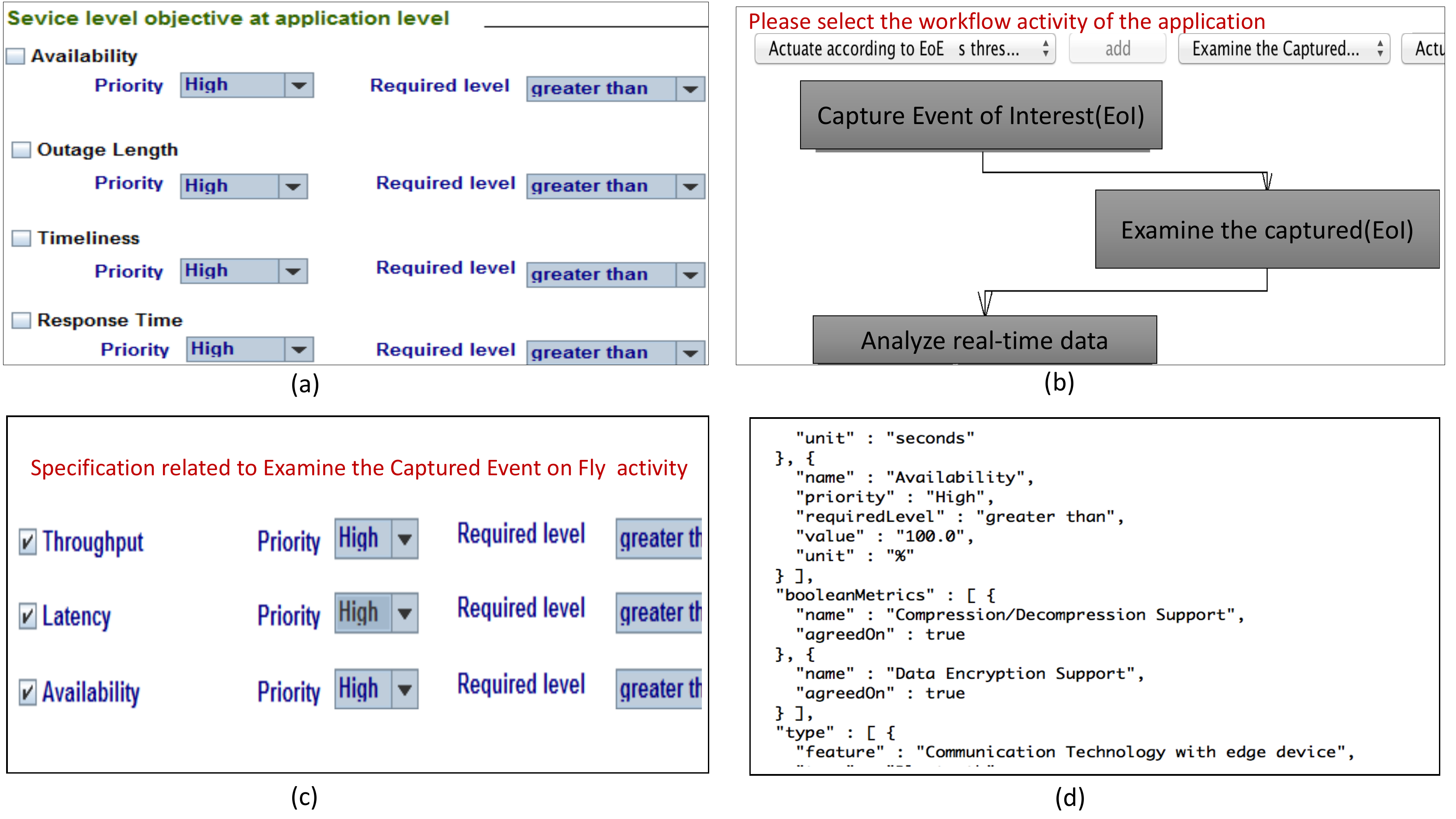} 
  \caption{ SLA specification toolkit: (a) Specify the SLOs of the Application, (b) Select the workflow activity based on the application scenario requirement, (c) Specify SLOs and configuration metrics related to each of the selected activities, and (d) Generate SLA document in JSON format.}
  \label{fig1} 
\end{figure}

\item \textit{Select the workflow activity based on the application scenario:} 
There is a predefined list of activities which are part of many IoT applications' workflow (e.g., capture event of interest; ingest data; analyze large-scale real-time data activity). The tool displays the predefined activity and then the user can select the ones that are included in their application workflow activities and connect them in a way that reflects the data flow of the application. Connecting the activity preserves the dependencies between activities for future work related to performance modelling. For example, the workflow activity of RHMS can consist of the following activities which can be connected, sequentially, in the same order as listed below (Figure 3-b):
1) Capture Event of Interest (EoI). 2) Examine captured EoI. 3) Ingest data activity. 4) Real-time Analysis activity. 5) Store structured data activity. The reason behind considering very standard and common activities is to increase the generality of the tool. 

\item 	\textit{Specify SLO and configuration metrics related to each of the selected activities:} 
The tool reads the SLO and configuration metrics schema from a predefined Excel file, which has the schema content of what to display for each activity.
Then, users can specify the required level/value of an SLO and the configuration metrics for each software/hardware component required to deliver the selected activities (Figure 3-c). In RHMS, for example, when the user selects a  \enquote{capturing event of interest} activity, the user can then specify the requirements at the IoT devices level, such as sampling rate, battery life and communication mechanism. 
\item \textit{Generate SLA document:} Based on what the user has specified in steps 1 and 3, the SLA document will be generated in a JSON format. The generated SLA can be used later on for different purposes, such as service provider discovery, SLA-based monitoring and SLA-based resource allocation. In RHMS, when the user presses \textit{finish}, a JSON document is generated based on what has been specified, which represents the SLA of the RHMS (Figure 3-d). The JSON document's specification is related to the SLOs at the application level followed by the specification related to each activity, which includes the SLOs and configuration related metrics for the required programming model (such as stream processing) and deployment layer.

\item  \textit{Store generated SLA  using a NoSQL database}: Since users can choose which metrics to specify by checking/unchecking the metrics, Therefore,generated SLA  has different schemas of JSON files being created, due to the heterogeneity of requirements from  different users. Therefore, each generated SLA JSON file is stored in a NoSQL database (MonogDB database). The reason behind storing required SLA metrics is for SLA compliance monitoring, which is one part of the SLA's life cycle.
\end{itemize}

\section{Conclusion and Future Work}\label{sec:conc}
We have presented a tool that supports end to end specification of QoS requirements within SLAs for IoT applications. The tool is used to simplify the process of capturing the requirements of IoT applications. We believe the tool effectively tackles the aforementioned challenges: 1) Specifying the requirements of an IoT application that has a multi-layered nature. The tool has provided a rich set of vocabularies to capture the requirements of each layer of the IoT architecture (IoT device, Edge layer, Cloud layer). 2) IoT applications have different SLO requirements, which vary from one application to another. Also the priority level of one SLO differs from one application to another. Therefore the tool allows the users to specify SLOs at the application level. 3) Different IoT applications have different workflow activities depending on each application use case scenario. Therefore, to overcome the varied requirements that arise from the heterogeneity of workflow activity, the tool allows the users to specify their workflow activity first and then use that to specify the requirements related to the hardware and software components of each selected activity. The output of the tool is an SLA specification in a machine-readable format (JSON format).

For future work, the SLA document will be used as an input to an SLA-based broker framework for discovering and negotiating with providers of IoT-related services. This is in order to find the best match of available providers who can deliver the service within the users' requirements. Additional work will investigate the implementation of a framework for monitoring that IoT applications meet the SLOs and QoS metrics specified within the SLA. The monitoring framework will be based on novel blockchain and smart contract technology~\cite{103}~\cite{104}. \\

\begin{IEEEkeywords}
\textit{Clouds and Edge Computing and Applications; Service Level Agreement;  SLA Specification; IoT.}\\
\end{IEEEkeywords}
\centering\textbf{BIOGRAPHIES}\\

\vskip -2\baselineskip plus -1fil
\begin{IEEEbiographynophoto}
{Awatif Alqahtani} 
has a BS and MS in Computer Science from King Saud University, Saudi Arabia. She is currently
working toward a Ph.D. in the School of Computing
Science at Newcastle University, UK. Her research interests includes Internet of Things, Big Data and service level.
Contact her at
a.alqahtani@newcastle.ac.uk
\end{IEEEbiographynophoto}
\vskip -2\baselineskip plus -1fil
\begin{IEEEbiographynophoto}
{Pankesh Patel} 
is working at Fraunhofer USA – Center for Experimental Software Engineering (CESE) as a Senior Research Scientist. His current focus is on implementation of Industry 4.0 techniques and methodologies in commercial environments. Prior to joining Fraunhofer, he served as a Research Scientist for the Industrial Software System group at ABB Corporate Research-India. 
He is frequently invited to speak as a panelist and keynote at both academic and commercial conferences. Pankesh obtained his Ph.D. from the University of Paris VI and the French National Institute for Research in Computer Science and Automation (INRIA) in Paris, France.

\end{IEEEbiographynophoto}
\vskip -2\baselineskip plus -1fil
\begin{IEEEbiographynophoto}
{Ellis Solaiman } 
is a Lecturer at the School of Computing,
Newcastle University. He previously received his Ph.D.
in Computing Science also from Newcastle University,
where he subsequently worked as a Research Associate
and Teaching Fellow. His research interests are mainly
in the areas of Dependability and Trust in Distributed
Systems such as the Cloud and the Internet of Things. He
is also interested in the automated monitoring of these
systems using technologies such as Smart Contracts. He
is a Fellow of the UK Higher Education Academy (FHEA)
since 2016.
\end{IEEEbiographynophoto}
\vskip -2\baselineskip plus -1fil
\begin{IEEEbiographynophoto}
{Rajiv Ranjan} 
is an Associate Professor (Reader) in Computing
Science at Newcastle University, United Kingdom.
Prior to that, he was a Senior Research and Julius Fellow
at CSIRO, Canberra, where he was working on projects
related to Cloud and big data computing. He has been conducting
leading research in the area of Cloud and big data
computing developing techniques. 
He has published about 110 papers that include 60+ journal papers. 
He serves on the editorial board of IEEE Transactions on Computers, IEEE
Transactions on Cloud Computing, IEEE Cloud Computing, and Future Generation Computer System Journals. He is one of the highly cited authors (top 0.09\%) in
computer science and software engineering worldwide (h-index=43, g-index=94,and 10,050+ google scholar citations).
\end{IEEEbiographynophoto}

\renewcommand\refname{\textbf{REFERENCES}}
\bibliographystyle{IEEEtran}
\bibliography{ref} 




\end{document}